\documentclass[conference]{IEEEtran}
\IEEEoverridecommandlockouts
\usepackage[T1]{fontenc}
\usepackage{times}
\usepackage[font=footnotesize,caption=false]{subfig} % preload with correct options...

\usepackage{cite}
\usepackage{amsmath,amssymb,amsfonts}
\usepackage{algorithmic}
\usepackage{graphicx}
\usepackage{textcomp}
\usepackage{xcolor}
\usepackage{booktabs}
\usepackage{algorithm}
\usepackage{bbm}
\usepackage{bm}
\usepackage{multirow}
\usepackage{tabularx}
\usepackage[utf8]{inputenc}
\usepackage{tikz}
\usepackage{siunitx}
\usepackage{cleveref}
\usepackage{algorithm, algorithmic}
\usepackage{makecell}

\usetikzlibrary{shapes,arrows}
\usetikzlibrary{plotmarks}
\usetikzlibrary{patterns}
\usetikzlibrary{arrows.meta}
\usetikzlibrary{positioning}
\usetikzlibrary{spy}

\newcommand{\set}{\mathcal}

\newcommand{\ex}[1]{\operatorname{\mathbb{E}}\!\left[#1\right]}       % Expected value
\newcommand{\var}[1]{\operatorname{Var}\!\left(#1\right)}     % Variance
\newcommand{\tsub}[1]{_{\text{#1}}}
\newcommand{\tsup}[1]{^{\text{#1}}}

\usepackage[nopostdot,nogroupskip,style=super,nonumberlist,acronym]{glossaries}

\def\BibTeX{{\rm B\kern-.05em{\sc i\kern-.025em b}\kern-.08em
    T\kern-.1667em\lower.7ex\hbox{E}\kern-.125emX}}
\newacronym{leo}{LEO}{Low Earth Orbit}
\newacronym{eo}{EO}{Earth Observation}
\newacronym{gs}{GS}{Ground Station}
\newacronym{flop}{FLOP}{Floating Point Operation}
\newacronym{bsp}{BSP}{Bulk-Synchronous Parallel}
\newacronym{rv}{RV}{Random Variable}
\newacronym{cdf}{CDF}{Cumulative Density Function}
\newacronym{oec}{OEC}{Orbital Edge Computing}
\newacronym{dnn}{DNN}{Deep Neural Network}
\newacronym{ai}{AI}{Artificial Intelligence}
\newacronym{e2e}{E2E}{End-to-End}
\newacronym{ntn}{NTN}{Non-Terrestrial Network}
\newacronym{fso}{FSO}{Free-Space Optical}
\newacronym{isl}{ISL}{Inter-Satellite Link}
\newacronym{awgn}{AWGN}{additive white Gaussian noise}
\newacronym{snr}{SNR}{Signal-to-Noise Ratio}
\newacronym{i.i.d.}{i.i.d.}{Independent and Identically Distributed}
\newacronym{fbl}{FBL}{Finite Blocklength}
\newacronym{arq}{ARQ}{Automatic Repeat Request}
\newacronym{pdf}{pdf}{Probability Density Function}
\newacronym{ul}{UL}{Uplink}
\newacronym{dl}{DL}{Downlink}
\newacronym{mgf}{MGF}{Moment Generating Function}
\newacronym{dt}{DT}{digital twin}
\newacronym{ue}{UE}{user}
\newacronym{mle}{MLE}{Maximum Likelihood Estimation}
\newacronym{prb}{PRB}{Physical Resource Block}
\newacronym{ofdma}{OFDMA}{orthogonal frequency-division multiple access}
\newacronym{ofdm}{OFDM}{orthogonal frequency-division multiplexing}

\begin{document}
\title{Statistical Analysis for Energy-Efficient Satellite Edge Computing with Latency Guarantees}
%\author{Author 1, Author 2 ...}

\author{\IEEEauthorblockN{Nicolai Dalsgaard Lyholm\IEEEauthorrefmark{1}, Beatriz Soret\IEEEauthorrefmark{2}\IEEEauthorrefmark{1}, Tijana Devaja\IEEEauthorrefmark{1},\\Thomas Gundgaard Mulvad\IEEEauthorrefmark{1}, Cedomir Stefanovic\IEEEauthorrefmark{1}, and Israel Leyva-Mayorga\IEEEauthorrefmark{1}}
\IEEEauthorblockA{\IEEEauthorrefmark{1} Department of Electronic Systems, Aalborg University, Denmark\\
\IEEEauthorrefmark{2} Telecommunications Research Institute, University of Málaga, Spain \\
Emails: \{ndly, bsa, tde, tgm, cs, ilm\}@es.aau.dk,  
} 
}

\maketitle

\begin{abstract}
Being able to provide latency guarantees for orbital edge computing applications through \gls{leo} satellite constellations is a major milestone for their integration into 5G and 6G networks. However, achieving this is fundamentally challenged by the inherent randomness in both communication and computing latency, driven by complex network dynamics, satellite motion, and hardware variability. 
In this paper, we perform a statistical analysis of the latency of satellite edge computing using representative computing hardware and an object detection algorithm running on a satellite image dataset. 
The resulting model captures the trade-off between data availability and estimation uncertainty, enabling data-driven optimization methods to meet latency targets with statistical guarantees while minimizing energy consumption.
%Our analysis enables the digital twinning of satellite edge computing applications, illustrating the trade-off between data availability and estimation uncertainty, and is complemented with data-driven optimization methods to achieve the application latency target with reliability guarantees and minimum energy consumption. 
Our results show that parametric estimation and quantile regression for the execution time of the image processing algorithms can be effectively combined with models for the communication latency to select an optimal GPU clock frequency.
This achieves a $95$\% probability of meeting a $500$\,ms end-to-end deadline while reducing energy consumption by more than $50$\% compared to a baseline that relies on a Chebyshev–Cantelli inequality to bound execution-time quantiles.
The proposed framework is generalizable across satellite edge computing workloads and hardware platforms.
\end{abstract}

\section{Introduction}\label{sec:introduction}
\gls{leo} satellite constellations are rapidly transforming \gls{eo} by enabling continuous capture of high-resolution imagery at a large scale. However, the volume of data generated on-board far exceeds the available downlink capacity, creating a fundamental bottleneck in traditional store-and-forward architectures \cite{yin_survey_2025}. This disparity has motivated the shift towards \gls{oec}, where computational resources are deployed and exploited directly on the satellite in order to process data closer to the point of acquisition, transmitting only the extracted semantic information to ground \cite{denby_orbital_2019}. By performing tasks such as object detection, cloud filtering, and scene classification on-board, \gls{oec} dramatically reduces the required \gls{dl} bandwidth enabling decision making for time-sensitive applications including maritime surveillance, disaster response, and environmental monitoring.

A key enabler of \gls{oec} is the advancements and increased availability of GPU-accelerated edge hardware capable of executing \gls{ai} algorithms within the strict size, weight, and power constraints of satellite platforms. Recent work has shown that such devices exhibit complex, workload-dependent performance characteristics: execution times vary with operating frequency, power mode and the computational demands of the inference task \cite{sk_performance_2023,SATNEX}. Understanding and accurately modeling these characteristics is essential for scheduling and resource management in orbital environments, where the energy budgets are tightly constrained and processing must be completed within strict timing windows dictated by the orbital geometry and the mission planning. The relevance of GPU-accelerated hardware in space is further underscored by NVIDIA's announcement at GTC 2026 that the Jetson Orin platform is being adopted for on-orbit \gls{ai} inference by satellite operators including Kepler communication and Sophia Space \cite{nvidia_gtc2026_space}, establishing Jetson Orin as a representative platform for \gls{oec} research.
\begin{figure}[t]
    \centering
    \includegraphics[width=\linewidth]{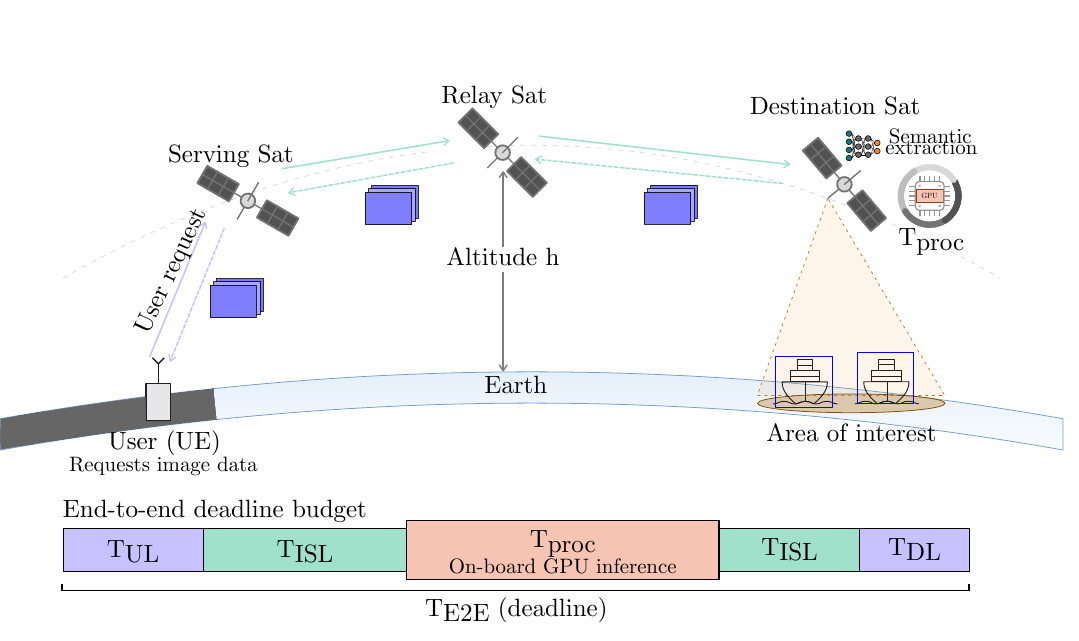}
    \caption{Satellite edge computing scenario and end-to-end latency budget $T\tsub{E2E}$, decomposed into uplink ($T\tsub{UL}$), inter-satellite ($T\tsub{ISL}$), on-board processing ($T\tsub{proc}$) and downlink ($T\tsub{DL}$) components.}
    \label{fig:scenario}
\end{figure}

In this paper, we consider a satellite edge computing scenario where users on ground request information contained in satellite images from a specific geographic area. The users send requests to their serving satellite, which relays the information to a destination satellite with visibility over the area of interest. The destination satellite captures the images and, based on our statistical model, chooses the best processing frequency to extract the semantic information from the image with minimum energy consumption while achieving statistical guarantees of meeting the latency constraints of the application. Our main contributions are:
\begin{enumerate}
    \item We present a statistical framework to characterize the per-image execution time of an object detection algorithm on representative on-board GPU hardware as a Gamma-distributed random variable with frequency-dependent shape and scale parameters.
    \item We formulate the selection of the GPU clock frequency as an energy-minimization problem under a quantile-based reliability constraint on the end-to-end latency, incorporating a finite-blocklength model of the satellite to user request transmission.
    \item We compare the performance of our framework against a benchmark that obtains a closed-form upper bound on the optimal frequency from the Chebyshev-Cantelli inequality without imposing a distributional assumption.
    \item We quantify the impact of sample size on the ability of the parametric model to yield reliable scheduling decisions, highlighting the tradeoff between model training effort and estimation uncertainty.
\end{enumerate}

\section{System model}\label{sec:system_model}
We consider the 5G \gls{ntn} scenario illustrated in Fig.~\ref{fig:scenario}, in which a \gls{ue} on ground requests information contained in satellite images from a specific geographical area. 
The request is transmitted to the serving satellite via the \gls{ul}, forwarded through \glspl{isl} to a destination satellite with visibility of the target area.
The destination satellite captures $N\tsub{img}$ images, processes them to extract semantic information (e.g. ship positions via object detection), and returns the result to the user through \glspl{isl} and the \gls{dl}. Both the request and the inference output are small data blocks exchanged over the network, while the computational load is concentrated onboard the satellite. The task is latency-constrained: the \gls{e2e} delivery must complete within a maximum latency $T\tsub{E2E}$. The satellite chooses an operating frequency for its on-board GPU computing hardware, which minimizes energy consumption while statistically fulfilling the latency constraint.

\subsection{Communication Model}
The satellites communicate with each other through \gls{fso} \glspl{isl} that operate at a fixed rate $R\tsub{isl}$~bps.
\gls{ul} and \gls{dl} communication between the user and the serving satellite occurs over an \gls{awgn} channel using radio-frequency links with frequency $f\tsub{comm}$~Hz. At a given time instant, the satellite-to-user links experience free-space path loss
\begin{equation}
    \set{L}(d) = \left(\dfrac{4\pi d \,f\tsub{comm}}{v_c}\right)^2,
\end{equation}
where  $v_c$ is the speed of light and $d$ is the distance between the user and the serving satellite. Note that the latter is dependent on the altitude of the serving satellite $h$ and its elevation angle at the specific time instant.
Next, let $G\tsub{sat}$ and $G\tsub{UE}$ denote the antenna gain of the satellite and the UE, respectively, and let $P_\ell$ be the transmission power for link $\ell\in\{\text{DL}, \text{UL}\}$. The \gls{snr} for communication between the UE and the satellite, separated by a distance $d$ is
\begin{equation}
    \gamma_\ell(d)=\frac{P_\ell G\tsub{sat} G\tsub{UE}}{\set{L}(d)\,\varphi\,\sigma_\ell^2 \Psi},
    \label{eq:snr}
\end{equation}
where $\varphi$ is the pointing loss, $\sigma_\ell^2$ is the noise power at the receiver and $\Psi$ is a log-normal shadow fading variable, i.e. $10\log_{10}(\Psi)\sim\mathcal{N}(0,\sigma\tsub{sf}^2)$.

Given that the request for satellite image data requires only a few bytes and the transmission power at the UEs is limited, we consider \gls{fbl} effects in the \gls{ul}, where the block error probability depends on the block length $n$, code rate $R$, and instantaneous \gls{snr} $\gamma$ \cite{5452208,Tijana24} as
% \td{In contrast to classical asymptotic communication models, we consider short-packet transmissions typical for edge computing and IoT applications, where the FBL regime must be taken into account. In this regime, the decoding error probability is no longer characterized by a sharp SINR threshold but depends explicitly on the blocklength $n$, code rate $R$, and instantaneous signal-to-noise ratio $\gamma$. The block error probability can be expressed as \cite{5452208}
\begin{equation}
    \epsilon(\gamma) = Q\!\left( \sqrt{\frac{n}{V(\gamma)}} \big(C(\gamma) - R\big) \right),
   \label{eq:FBL_error} 
\end{equation}
with $C(\gamma)=\log_2(1+\gamma)$ being the Shannon capacity and $\mathit{V(\gamma)}=\gamma\frac{\gamma+2}{(1+\gamma)^2}\log _2^2( e)$ being the channel dispersion. An \gls{arq} mechanism is implemented to retransmit failed packets and, hence, provide reliable data delivery despite finite blocklength effects. Due to a considerably  higher transmit power at the satellite $P\tsub{DL}\gg P\tsub{UL}$, NACK transmissions are assumed to be error free and the number of transmission attempts follows a geometric distribution with success probability $1-\epsilon(\gamma)$. 

The uplink blocklength $n$ (in channel uses) required by the finite blocklength model is mapped onto the \gls{ofdm} resource grid of 5G \gls{ntn}, where $N\tsub{SC}$ channel uses can be conveyed simultaneously within each symbol duration $T\tsub{symb}$. Therefore, the \gls{ul} delay per transmission attempt, including propagation, is %$T\tsub{tx} = n/R+d/v_c$ 
$T\tsub{tx} = T\tsub{symb}\left\lceil n/N\tsub{SC}\right\rceil+d/v_c$.  Hence, the uplink delay distribution including retransmissions and NACK delays is
\begin{IEEEeqnarray}{rCl}
\Pr\big(T_{\text{UL}}(R,d) &=& T_{\text{tx}} + x(T_{\text{tx}} + T_{\text{NACK}})\big) \nonumber \\
&=& (1 - \epsilon(\gamma_{\text{UL}}(d))) \epsilon(\gamma_{\text{UL}}(d))^x,
\end{IEEEeqnarray}
where $x$ is the number of failed attempts and $T\tsub{NACK}$ is the fixed NACK delay. The expected uplink delay is
\begin{equation}
    \ex{T\tsub{UL}(R,d)} = T\tsub{tx}+\frac{\left(T\tsub{tx}+T\tsub{NACK}\right)\epsilon(\gamma\tsub{UL}(d))}{1-\epsilon(\gamma\tsub{UL}(d))} %\frac{T_{\mathrm{tx}}}{1-\epsilon(\gamma)}.
    \label{ul_delay}
\end{equation}
While the use of short block lengths may result in \gls{fbl} effects in the \gls{dl}, the error probability is negligible due to the significantly higher transmission power at the satellites, assuming the same block length $n$ as in the \gls{ul}. Hence, $T\tsub{DL}= T\tsub{symb}\left\lceil n/N\tsub{SC}\right\rceil+d\tsub{DL}/v_c$ is treated as deterministic.
% While the use of short block lengths $n\tsub{DL}$ for transmitting the inference result in the \gls{dl} may result in \gls{fbl} effects, as with NACK transmissions, the \gls{dl} error probability is negligible due to the significantly higher transmission power at the satellites. Hence, $T\tsub{DL}= T\tsub{symb}\left\lceil n/N\tsub{SC}\right\rceil+d\tsub{DL}/v_c$ is treated as deterministic, where we assume the same blocklength $n$ as in the \gls{ul}.

Retransmissions are also neglected during inter-satellite communication, such that the delay at each \gls{isl} is composed of a transmission and a propagation component. For a block of size $n$ bits transmitted over a path $\mathcal{P}$, where the path itself determines the number of hops, and where the inter-satellite distance of each link $e\in\mathcal{P}$ is denoted as $d\tsub{ISL}(e)$, the round-trip \gls{isl} latency is given by
\begin{equation}
    T\tsub{ISL} = \sum_{e\in\mathcal{P}}2\,T\tsub{symb}\tsup{ISL}  \left\lceil\frac{n}{N\tsub{SC}\tsup{ISL}}\right\rceil + \frac{2\,d\tsub{ISL}(e)}{v_c},%\frac{n}{R\tsub{ISL}} + \frac{n\tsub{proc}}{R\tsub{ISL}} + \frac{2d\tsub{ISL}(e)}{v_c},
\end{equation}
where $T\tsub{symb}\tsup{ISL}$ and $N\tsub{SC}\tsup{ISL}$ are the duration of a symbol and the number of subcarriers in the \gls{isl}, respectively.

\subsection{Processing Model}

Each processing node in the constellation is characterized by: the number of processing cores $N\tsub{cores}$, the maximum operating frequency $f\tsub{max}$, the \glspl{flop} per clock cycle per core $N\tsub{FLOPs}$, and the maximum power consumption $P\tsub{max}$. The power consumption at operating frequency $f_p$ follows the cubic scaling model
\begin{equation}
    P(f_p)=P\tsub{max}\left(\frac{f_p}{f\tsub{max}}\right)^3.
\end{equation}
The execution time for processing a single image at frequency $f_p$ is modeled by the \gls{bsp} model as
\begin{equation}\label{eq:BSP}
    T\tsub{exec}(f_p)=\frac{CW}{N\tsub{cores}N\tsub{FLOPs}f_p}+T\tsub{sync},
\end{equation}
where $W$ is the algorithmic work in \glspl{flop}, $C\geq 1$ is a \gls{rv} capturing processing inefficiency, and $T\tsub{sync}$ captures synchronization overhead within the processing unit.
% Let $W$ denote the algorithmic work, defined as the number of \glspl{flop} required to execute the algorithm on a single image. The execution time for processing a single image at processing node $p$ at frequency $f_p$ is modeled by the \gls{bsp} model as
% \begin{equation}\label{eq:BSP}
% T\tsub{proc}(f_p)=T\tsub{work}(f_p,C) + T\tsub{sync}
% \end{equation}
% where $T\tsub{sync}$ captures the delay due to communication and synchronization delay among processing cores and the time to execute the algorithmic work is
% \begin{equation}
%     T\tsub{work}(f_p,C) = \dfrac{C W}{N\tsub{cores} N\tsub{FLOPs} f_p},
% \end{equation}
% with $C\geq1$ being a \gls{rv} which captures the processing inefficiency of the algorithm due to overhead from task allocation or memory management.

\subsection{Statistical Characterization of Execution Time} \label{subsec:Stat_characterization}
Taking the expectation of the \gls{bsp} model \eqref{eq:BSP}, the mean execution time is given as
\begin{IEEEeqnarray}{rCl}
\label{eq:mean_exec_bsp}
    \ex{T\tsub{exec}(f_p)} = \dfrac{\mu_C W}{N\tsub{cores} N\tsub{FLOPs} f_p}+\mu\tsub{sync},
\end{IEEEeqnarray}
where $\mu_C=\ex{C}$ and $\mu\tsub{sync}=\ex{T\tsub{sync}}$. The least squares estimators for $\mu_C$ and $\mu\tsub{sync}$ for the case where a sample of size $N_s$ for $T_{\text{exec}}\left(f_p\right)$ is available for multiple values of $f_p$ are, respectively
\begin{align}
    \hat{\mu}_\text{sync} & =\frac{1}{N_s}\sum_{i=1}^{N_s}T_i-\frac{\hat{\mu}_C}{N_s}\sum_{i=1}^{N_s}\frac{1}{f_p^i} \\
    \hat{\mu}_C & = \frac{\displaystyle\frac{1}{N_s}\left(\sum_{i=1}^{N_s}T_i\right)\sum_{j=1}^{N_s}\frac{1}{f_p^j}-\sum_{i=1}^{N_s}\frac{T_i}{f_p^i}}{\displaystyle\frac{1}{N_s}\left(\sum_{i=1}^{N_s}\frac{1}{f_p^i}\right)^2-\sum_{i=1}^{N_s}\frac{1}{\left(f_p^i\right)^2}},
\end{align}
where $T_i$ is the execution time (i.e., outcome) of the $i$-th image (i.e., experiment) using processing frequency $f_p^i$. 

Beyond the mean, the full distribution of the execution time is essential for providing scheduling guarantees. Throughout this paper, we model the per-image execution time for algorithms running on GPU hardware as a Gamma-distributed \gls{rv}
\begin{equation}
    T\tsub{exec}(f_p)\sim \text{Gamma}(\alpha(f_p),\theta(f_p)),
\end{equation}
with mean $\ex{T\tsub{exec}(f_p)} = \alpha(f_p)\,\theta(f_p)$ and variance $\var{T\tsub{exec}(f_p)} = \alpha(f_p)\,\theta(f_p)^2$. The goodness of fit of the Gamma distribution to the execution time with the considered hardware and algorithm was the highest among a wide range of candidate distributions, reaching a Kolmogorov-Smirnov statistic of 0.05, further supported by achieving the best AIC and BIC statistics among the candidate distributions. These findings are in line with observations in the literature \cite{Lorch01,suman_analysis_2022}.

Building on this, the frequency dependent shape and scale parameters are estimated using polynomial regression
\begin{equation}
    \hat{\alpha}(f_p) = \sum_{i=0}^L c_{\alpha,i}f_p^i, \quad \hat{\theta}(f_p)=\sum_{i=0}^L c_{\theta,i}f_p^i,
\end{equation}
where $L$ is the polynomial order and $c_{\alpha,i}$ and $c_{\theta,i}$ are the fitted coefficients for the shape and scale respectively.

When processing $N\tsub{img}$ images sequentially at a constant operating frequency $f_p$, the total processing time is the sum of $N\tsub{img}$ \gls{i.i.d.} Gamma \glspl{rv}. Using the \gls{mgf}, we have
\begin{IEEEeqnarray}{rCl}
    M_{T_{\text{exec}}}(t) &=& \Pi_{i=1}^{N\tsub{img}} \left(1-\theta(f_p) t \right)^{-\alpha(f_p)} \nonumber\\
    &=& \left(1-\theta(f_p) t \right)^{-N\tsub{img}\alpha(f_p)},
\end{IEEEeqnarray}
which is the \gls{mgf} of a Gamma distribution with shape $N\tsub{img}\alpha(f_p)$ and scale $\theta(f_p)$. Therefore,
\begin{IEEEeqnarray}{rCl}
    T_{\text{exec}}(f_p,N\tsub{img}) = &\sum_{i=1}^{N\tsub{img}}&T_{\text{exec},i}(f_p) \nonumber \\
    &\sim& \text{Gamma}\left(N\tsub{img}\alpha(f_p),\theta(f_p)\right).
\end{IEEEeqnarray}

% \subsubsection{Characterization of Frequency-Dependent Parameters}
% The Gamma parameters $\alpha(f_p)$ and $\theta(f_p)$ are obtained empirically by fitting execution time measurements collected at multiple operating frequencies. We model their frequency dependence using polynomial regression as
% \begin{equation}
%     \alpha(f_p) = \sum_{i=0}^L c_{\alpha,i}f_p^i, \quad \theta(f_p)=\sum_{i=0}^L c_{\theta,i}f_p^i,
% \end{equation}
% where $L$ is the polynomial order and $c_{\alpha,i}$ and $c_{\theta,i}$ are the fitted coefficients for processing node $p$.
% \subsection{Scheduling with Quantile-Based Constraints}

%\subsection{E2E Latency and Optimization}
\begin{figure*}[t!]
    \centering
    \includegraphics[width=\linewidth]{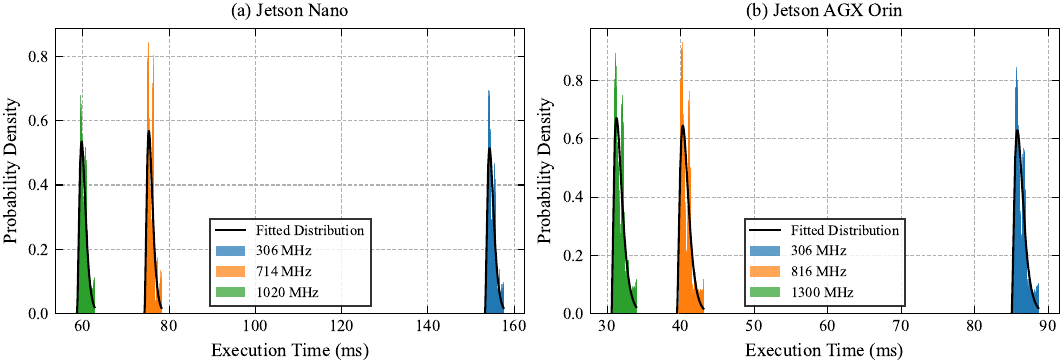}
    \caption{Empirical execution time distribution for (a) NVIDIA Jetson Nano and (b) Jetson AGX Orin at selected GPU operating frequencies. The solid black lines represent the fitted Gamma distributions.}
    \label{fig:Distribution_fit}
\end{figure*}

\subsection{Optimization}

The \gls{e2e} latency consists of the communication times for the \gls{ul}, the \gls{isl}, and the \gls{dl}, and the processing time. Once the data reaches the satellite responsible for computation, the communication delays, originally modeled as random variables, are realized as deterministic values. Then, the available processing time budget is written as
\begin{equation}
    T\tsub{proc}=T\tsub{E2E}-T\tsub{UL}-T\tsub{ISL}-T\tsub{DL},
\end{equation}
where $T\tsub{E2E}$ is the total \gls{e2e} time budget. 

%\nl{The \gls{e2e} latency comprises $T\tsub{E2E}=T\tsub{UL}+T\tsub{ISL}+T\tsub{proc}+T\tsub{DL}$. The scheduling decision is made after the user request arrives at the destination satellite, at which point $T\tsub{UL}$ has been realized and $T\tsub{ISL}$, $T\tsub{DL}$ are deterministic. The remaining processing budget is
%\begin{equation}
%    T\tsub{rem}=T\tsub{max}-T\tsub{UL}-T\tsub{ISL}-T\tsub{DL}
%\end{equation}}
%\ilm{While it is possible to calculate the minimum frequency that ensures the mean total processing time meets $T\tsub{proc}$ from the BSP model\,\cite{SATNEX}, this approach would not be suitable for providing statistical guarantees.}
%Using the BSP mean, the minimum frequency ensuring the mean total processing time meets the deadline is \cite{SATNEX}
%\begin{equation}
%    f^*_p= \dfrac{\mu_C W}{N\tsub{cores}N\tsub{FLOPs}\left(T\tsub{proc}-N\tsub{img}\mu\tsub{sync} \right)}.
%\end{equation}
%However, 
%this would result in an excessively high probability of exceeding the disregard stochastic variability and 
%provide no statistical guarantees on the probability on meeting the deadline. 
%We instead impose a reliability constraint:
Note that, since $P(f_p)$ is increasing monotonically, minimizing the operating frequency is equivalent to minimizing the energy consumption.
Therefore, the optimal processing frequency for a latency-constrained application that requires a reliability $\rho\tsub{th}$ is the minimum operating frequency such that the processing completes within $T\tsub{proc}$ with a probability of at least $\rho\tsub{th}$. Namely, for a given $N\tsub{img}\,\alpha(f_p),\theta(f_p)$, the optimal processing frequency is
\begin{equation}\label{eq:opt}
\begin{aligned}
f_p^* \triangleq\min%_{f_p \in [f_\mathrm{min}, f_\mathrm{max}]}
&\quad f_p\in [f_\mathrm{min}, f_\mathrm{max}] \\
\text{s.t.}&\; F_\mathrm{Gamma}\big(T_\mathrm{proc};N\tsub{img}\,\alpha(f_p),\theta(f_p)\big) \geq \rho\tsub{th}.
\end{aligned}
\end{equation}
%which is solved numerically over the discrete set of available frequencies on the computing hardware. Equivalently, the constraint requires $Q_{\rho\tsub{th}}(T\tsub{exec}(f_p,N\tsub{img}))\leq T\tsub{proc}$, where $Q_{\rho\tsub{th}}(\cdot)$ denotes the $\rho\tsub{th}$-quantile. 
\subsection{Benchmark}
As a benchmark that uses only the first two moments of the execution time distribution, the Chebyshev-Cantelli inequality provides a one-sided bound on the probability of the total processing time exceeding the deadline, given by
\begin{multline}
\Pr\!\left(T\tsub{exec}(f_p,N\tsub{img}) \geq t\right) \\
\leq \frac{N\tsub{img}\,\sigma^2(f_p)}{N\tsub{img}\,\sigma^2(f_p) + \big(t - N\tsub{img}\,\ex{T\tsub{exec}(f_p)}\big)^2}.
\end{multline}
valid for $t > N\tsub{img}\ex{T\tsub{exec}(f_p,N\tsub{img})}$, where $\sigma^2(f_p)=\var{T\tsub{exec}(f_p,N\tsub{img})}$. Imposing the reliability requirement that the right-hand side is at most $1-\rho\tsub{th}$ and substituting the \gls{bsp} mean from \eqref{eq:mean_exec_bsp} yields an upper bound on the optimal operating frequency
\begin{multline}
f^{*}_p \leq f\tsup{upper}_p(N\tsub{img}) \triangleq \frac{N\tsub{img}\mu_C W}{N\tsub{cores} N\tsub{FLOPs}} \\
\cdot \frac{1}{t - N\tsub{img}\mu\tsub{sync} - \sigma(f_p)\sqrt{N\tsub{img}\rho\tsub{th}/(1-\rho\tsub{th})}}.
\end{multline}

\subsection{Learning the Computation Time Distribution}
The optimization in \cref{eq:opt} requires knowledge of the Gamma parameters $\alpha(f_p)$ and $\theta(f_p)$, which must be learned empirically. In practice, a newly deployed satellite cannot afford to conduct an exhaustive sampling campaign before becoming operational. We study how scheduling reliability degrades when the Gamma parameters are estimated from a limited sample size.

Let $\mathcal{S}$ denote the full set of maritime images in the dataset. To emulate the situation in which only a limited number of execution time observations are available for parameter fitting, we draw a subset $\mathcal{S}_k$ of size $|\mathcal{S}_k|=N_s$ from $\mathcal{S}$ with replacement. Sampling with replacement allows $N_s>|\mathcal{S}|$, with repeated draws of the same image yielding distinct samples that capture the hardware's variability between executions. Each image in $\mathcal{S}_k$ is executed once per available operating frequency, yielding $N_s$ execution time samples at each frequency. From these samples the Gamma parameters are estimated via \gls{mle}, producing a subset characterization $(\hat{\alpha}_k(f_p), \hat{\theta}_k(f_p))$. To capture the variability of this characterization across different sample realizations, we repeat the procedure $K$ times for each sample size $N_s$, generating $K$ independent subset characterizations and consequently, $K$ frequency selections $\hat{f}^*_{p,k}$ obtained by solving \eqref{eq:opt} under each subset.
Each estimated frequency $\hat{f}^*_{p,k}$ is evaluated against the ground-truth Gamma distribution to compute the probability of missing the deadline due to a poor estimate
\begin{equation}\label{eq:miss_prob}
    P^{(k)}\tsub{miss} = 1-F_{T\tsub{exec}}(T\tsub{proc}; N\tsub{img} \alpha(\hat{f}^*_{p,k}),\theta(\hat{f}^*_{p,k})).
\end{equation}
If $P^{(k)}\tsub{miss}\approx1-\rho\tsub{th}$, the subset characterization is reliable. On the other hand, if $P^{(k)}\tsub{miss}\gg 1-\rho\tsub{th}$ the subset underestimated the execution time tail and operates with false confidence. Vice versa, if $P^{(k)}\tsub{miss}\ll 1-\rho\tsub{th}$, the subset is overly conservative, selecting a too high operating frequency and wasting energy.

\section{Results}

We evaluate the proposed framework in two parts. First, we validate the Gamma distribution model and assess its increase in accuracy as the sample size increases. Then, we compare the energy consumption of the Chebyshev-Cantelli baseline, imposing no distributional assumption on the execution time, against the Gamma quantile-based scheduling under varying workload and link conditions.

\subsection{Experimental Setup}
We validate the proposed framework using YOLOv8m \cite{yolov8_ultralytics} for ship detection in satellite imagery, executed on two NVIDIA Jetson platforms: the Orin Nano and the AGX Orin. While the framework is algorithm- and platform agnostic, this combination reflects two tiers of realistic \gls{oec} deployment scenarios \cite{nvidia_gtc2026_space}. We use the Ships-Google-Earth dataset, which is representative of the target Earth observation application and contains 56 satellite images of maritime scenes with varying ship densities \cite{ships-google-earth_dataset}. For the ground truth characterization, each image is executed 1000 times at every GPU frequency. The Gamma parameters $\alpha(f_p),\theta(f_p)$ are fitted via \gls{mle} and their frequency dependence is modeled by third-order polynomial regression. Communication, computing and simulation parameters are summarized in \Cref{tab:sim_params}. Throughout the evaluation, we set the reliability requirement $\rho\tsub{th}=0.95$ and the \gls{e2e} deadline to $T\tsub{E2E}=\SI{500}{ms}$ unless stated otherwise.

\subsection{Execution Time Characterization}
Fig.~\ref{fig:Distribution_fit} shows the empirical execution time distributions at three selected frequencies, overlaid with the fitted Gamma \glspl{pdf}. The Gamma model captures the shape, location and spread of the execution time samples across the full frequency range, confirming its suitability for characterizing inference execution times on GPU edge hardware. The frequency-dependent parameters $\alpha(f_p)$ and $\theta(f_p)$ are well described by the polynomial model, achieving a coefficient of determination $> 0.99$ for $\alpha$ and $>0.95$ for $\theta$ on both platforms.

\begin{figure}[t]
    \centering
     \includegraphics[width=\linewidth]{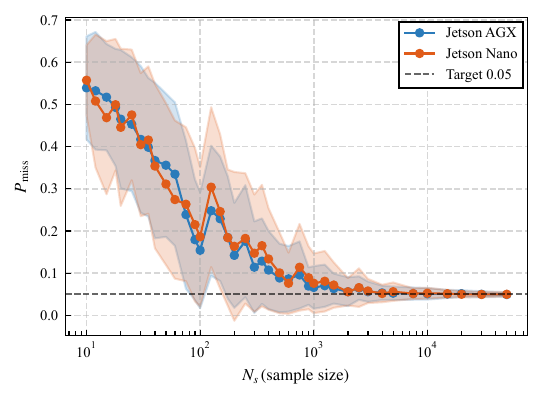}
    \caption{Deadline miss probability $P\tsub{miss}$ versus sample size $N_s$. Shaded regions show variation across the $K$ subsets for any choice of $N_s$.}
    \label{fig:subset_miss_probability}
\end{figure}

A key practical question is the sample size needed to learn a reliable characterization of the execution time. Fig~\ref{fig:subset_miss_probability} evaluates this by plotting the deadline miss probability $P\tsub{miss}^{(k)}$ as a function of the sample size $N_s$ used to fit the Gamma parameters. For each $N_s$, $K=100$ subsets are drawn with replacement, each producing an independent characterization and frequency selection via \eqref{eq:opt}. After this, the miss probability is evaluated against the ground truth distribution. At small $N_s$, the subset characterizations systematically underestimate the execution time tail, leading to too low frequency selections and miss probabilities significantly above the $1-\rho\tsub{th}=0.05$ target. As $N_s$ increases, the mean converges towards the target and the spread across subsets narrows. Beyond $N_s=1000$, the miss probability stabilizes within a small margin of the design target for both platforms, indicating that reliable scheduling can be achieved with a modest number of executions of the algorithm.

% \begin{figure}[H]
%     \centering
%     \includegraphics[width=\linewidth]{figures/gt_poly_fits.pdf}
%     \caption{\nl{Shows the empirical frequency dependent Gamma parameters (for the ground truth distribution with many samples), along with the polynomial fit. }}
%     \label{fig:Polyfits}
% \end{figure}

\subsection{E2E Optimization}
\begin{figure}
    \centering
    \includegraphics[width=\linewidth]{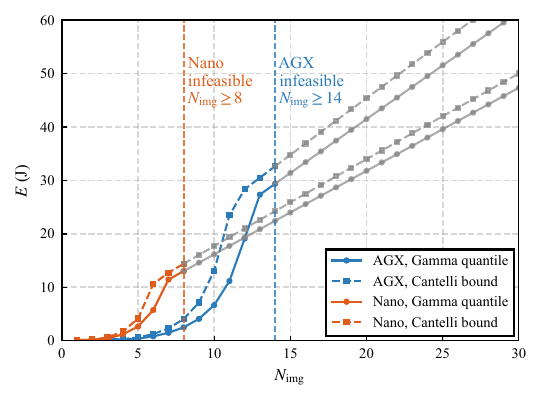}
    \caption{Processing energy $E$ versus number of requested images $N\tsub{img}$ for the proposed Gamma quantile-based scheduling and the Chebyshev-Cantelli baseline, under error-free communication.}
    \label{fig:Energy_vs_n}
\end{figure}

Fig~\ref{fig:Energy_vs_n} compares the processing energy consumption resulting from Gamma quantile-based scheduling against the Chebyshev-Cantelli baseline as a function of the number of requested images $N\tsub{img}$. 
We focus on the impact of processing time on the scheduling decision. To this end, the evaluation is conducted under conditions where $\epsilon(\gamma\tsub{UL}) \approx 0$, corresponding to high elevation angles, such that uplink ARQ retransmissions are negligible.

As $N\tsub{img}$ increases, the image processing time budget $T\tsub{proc}$ shrinks, requiring higher operating frequencies and thus greater energy consumption. The Gamma quantile-based method consistently selects a lower operating frequency than the Cantelli bound, since it exploits the full distributional knowledge, whereas the Cantelli bound is limited to the first two moments. The gap between the two methods increases as $N\tsub{img}$ grows and the deadline becomes tighter, highlighting the importance of distribution-aware scheduling under stringent deadlines. Beyond $N\tsub{img}=8$ for the Nano and $N\tsub{img}=14$ for the AGX, even the maximum operating frequency cannot satisfy the reliability constraint and the deadline becomes infeasible.
In practice, the available processing budget $T\tsub{proc}$ depends on the communication delays, which in turn depend on the satellite to user geometry. At low elevation angles the increased path-loss raises the \gls{ul} error probability $\epsilon(\gamma\tsub{UL})$, leading to more ARQ retransmissions and a larger expected \gls{ul} delay, thereby further reducing the time budget available for processing. 

\begin{figure}
    \centering
    \includegraphics[width=\linewidth]{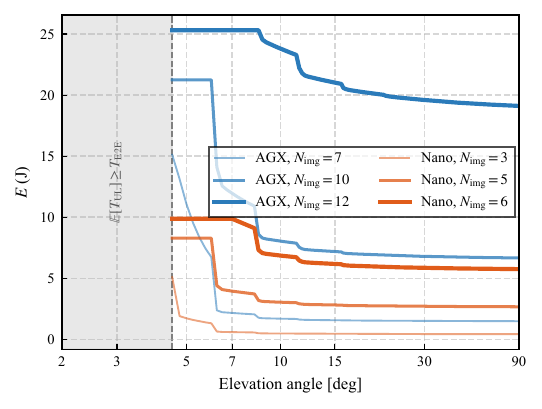}
    \caption{Processing energy $E$ versus satellite elevation angle for selected feasible values of $N\tsub{img}$ on both platforms. The shaded region marks elevations at which $\ex{T\tsub{UL}}\geq T\tsub{E2E}$, such that the task becomes infeasible.}
    \label{fig:energy_vs_elevation}
\end{figure}

Fig~\ref{fig:energy_vs_elevation}, evaluates this effect for both platforms at selected feasible values of $N\tsub{img}$, sweeping the elevation angle from near-zenith to near-horizon. At high elevation angles, few ARQ retransmissions occur and the energy consumption is dominated by the processing workload, matching the behavior observed in Fig~\ref{fig:Energy_vs_n}. As the elevation angle and $\gamma\tsub{UL}$ decrease, the \gls{ul} error probability $\epsilon(\gamma\tsub{UL})$ increases, causing more ARQ retransmission and a larger expected $T\tsub{UL}$. This reduces the processing budget, forcing the satellite to increase its operating frequency. Below a critical elevation angle (indicated by the shaded region), the expected UL delay consumes the entire \gls{e2e} deadline, making the task infeasible regardless of the processing strategy.

% \begin{figure}[H]
%     \centering
%     \includegraphics[width=\linewidth]{figures/subset_convergence.pdf}
%     \caption{\nl{Shows the convergence of subset characterizations towards the ground truth as the number of benchmark images increases. Shaded area indicates the best and worst performing subset at that choice of N. We will probably only keep one of these plots, since they tell the same story, one in terms of freqency selection error, the other in deadline miss probability.}}
%     \label{fig:subset_convergence}
% \end{figure}

\section{Conclusion}
This paper presented a statistical framework that enables the provision of reliability guarantees on the end-to-end latency of satellite edge computing applications while minimizing energy consumption. Our framework was validated on an Earth observation application with satellite image dataset and the YOLOv8m object detection algorithm executed on the NVIDIA Jetson Orin Nano and AGX platforms. In this scenario, the execution time was characterized as a Gamma-distributed random variable with frequency-dependent shape and scale parameters. %, enabling a quantile-based formulation of the energy-minimizing scheduling problem under latency deadline and reliability constraints. 
%We compared our approach, which leverages the full execution-time distribution, with a Chebyshev–Cantelli inequality baseline that relies only on the first two moments. 
Our method reduces the processing energy consumption by more than $50\%$ when compared to the selected benchmark, based on the Chebyshev–Cantelli inequality, while meeting a $95\%$ reliability target on a $500\,\text{ms}$ end-to-end deadline, with the gap widening as the deadline tightens. Furthermore, we showed that reliable parameter estimates can be obtained from a relatively small sample size of approximately $1000$, making the characterization practical to learn on-board.
Future work will focus on the extension and integration of our statistical framework into a \gls{dt} framework to enable continuous model adaptation and constellation-wide multi-service integration and orchestration.
%able to continuously refine its statistical models, not only for known algorithms but also for unseen workloads. 
%As seen in Fig.~\ref{fig:subset_miss_probability}, estimation accuracy improves with additional samples. Consequently, a DT that updates itself during operation would improve predictions over time.
%We envision this could enable constellation resource orchestration decisions, such as offloading tasks to neighboring satellites when deadlines cannot be met locally, or selecting the most energy-efficient node for execution.

\begin{table}[t]
\renewcommand{\arraystretch}{1.1}
\centering
\caption{Simulation parameters}
\begin{tabular}{@{}ll@{}ccc@{}}
    \toprule \multicolumn{2}{@{}l}{\textbf{Parameter}} & \textbf{Symbol} & \multicolumn{2}{c}{\textbf{Setting}} \\\midrule 
    \multicolumn{2}{@{}l}{\textbf{Communication}}\\
    \phantom{\,} &Carrier frequency [GHz] & $f\tsub{comm}$ & \multicolumn{2}{c}{$2$} \\
    &Bandwidth [kHz]  & $B$ & \multicolumn{2}{c}{$180$} \\
    & Satellite antenna gain [dBi] & $G\tsub{sat}$ & \multicolumn{2}{c}{$30$}\\
     &UE antenna gain [dBi] & $G\tsub{UE}$& \multicolumn{2}{c}{$0$}\\
    &Total number of satellites & $N\tsub{sats}$ & \multicolumn{2}{c}{$12$} \\
    & Number of ISL hops & $N\tsub{hops}$ & \multicolumn{2}{c}{$4$}\\
     &Altitude of deployment [km] & $h$  & \multicolumn{2}{c}{$600$}\\
     &Downlink transmission power [W] &$P\tsub{DL}$ & \multicolumn{2}{c}{$75$} \\ 
     &Uplink transmission power [W] &$P\tsub{UL}$ & \multicolumn{2}{c}{$0.2$} \\ 
     &Pointing loss [dB]  & $\varphi_\text{dB}$ & \multicolumn{2}{c}{$0.3$} \\
    &Noise spectral density [dBm/Hz] & $N_{0,\text{dB}}$ & \multicolumn{2}{c}{$-176.31$}\\
    &Shadow fading std.  [dB] & $\sigma\tsub{sf}$ & \multicolumn{2}{c}{$4$}\\
    & Number of subcarriers & $N\tsub{sc}$ & \multicolumn{2}{c}{$12$}\\
    & Number of OFDM symbols & $N\tsub{sym}$ & \multicolumn{2}{c}{$14$}\\
    & Subcarrier spacing [Hz] & $\text{SCS}$ & \multicolumn{2}{c}{$15\text{e}3$}\\
    & Number of slots allocated & $N\tsub{slots}$ & \multicolumn{2}{c}{$4$}\\
    & Payload size [bits] & $D$ & \multicolumn{2}{c}{$1500$}\\
    & Number of channel uses & $n$ & \multicolumn{2}{c}{672}\\
    & Transmission rate [bpcu] & $R$ & \multicolumn{2}{c}{$2.23$}
    \\\midrule
    \multicolumn{3}{@{}l}{\textbf{Computing with NVIDIA Jetson Orin}}&\textbf{Nano} & \textbf{AGX}\\
    &Number of cores [W] &$N_\text{cores}$ & 1024 & 2048\\
    & Max.\ power consumption [W] &$P_\text{max}$  &  25 & 60\\
& Max.\ clock frequency [GHz]&$f_\text{max}$  &  1.02 & 1.3\\
& Number of FLOPs per Hz &$N_\text{FLOPs}$ &  2 & 2\\
& Processing inefficiency &$\mu_C$ &   1.071 & 1.122\\
& Mean synchronization time [ms] &$\mu_\text{sync}$  &  17.48 & 14.14\\
& \makecell[l]{Mean execution time\\ at max. frequency [ms]} & $\ex{T\tsub{exec}(f_\text{max}^{(p)})}$  &$61.19$&$32.63$\\
    \bottomrule
    \end{tabular}
    \label{tab:sim_params}
\end{table}

\section{Acknowledgement}
 This work was supported, in part, by the Velux Foundation, Denmark, through the Villum Investigator Grant WATER, nr. 37793 and by Danmarks Frie Forskningsfond (DFF) Project “3D-Twin” under Grant 10.46540/4264-00153B. The work of B. Soret is partially supported by the Spanish Ministerio de Ciencia e Innovación under grant PID2022-136269OB-I00.

\bibliographystyle{IEEEtran}
\bibliography{bib.bib}

\end{document}